# Observation of Spin Wave Soliton Fractals in Magnetic Film Active Feedback Rings

Mingzhong Wu,[1] Boris A. Kalinikos,[1,2] Lincoln D. Carr,[3] and Carl E. Patton[1]

[1]*Department of Physics, Colorado State University, Fort Collins, Colorado 80523, USA*
[2]*St. Petersburg Electrotechnical University, 197376, St. Petersburg, Russia*
[3]*Department of Physics, Colorado School of Mines, Golden, Colorado 80401, USA*

The manifestation of fractals in soliton dynamics has been observed for the first time. The experiment utilized self-generated spin wave envelope solitons in a magnetic film based active feedback ring. At high ring gain, the soliton that circulates in the ring breathes in a fractal pattern. The corresponding power frequency spectrum shows a comb structure, with each peak in the comb having its own comb, and so on, to finer and finer scales.

Chaos, solitons, and fractals are key topics in the field of nonlinear science. All three have attracted intense interest over a wide range of disciplines, including mathematics, physics, biology, and engineering, to name a few [1-6]. In general terms, chaos involves nonlinear motion that occurs in deterministic dynamical systems but has the appearance of randomness. Solitons are localized large-amplitude pulse excitations that can occur in nonlinear dispersive systems. While chaotic motion appears irregular, soliton pulses are well defined and stable. They can travel without a change in shape and survive collisions. A fractal is a complex geometrical object made of parts that are similar to the whole in some way. There are regular fractals and random fractals. Regular fractals appear identical at different scales, while random fractals display statistic self-similarity.

While chaos and fractals have no direct relation *per se*, fractal structures very often appear in the phase space of chaos. The Hénon attractor, for example, has a Cantor set like fractal pattern. In view of the appearance of fractals in chaotic dynamics, it is also natural to hypothesize the appearance of fractals in soliton processes. Recent theoretical work indicates that fractal phenomena should exist in soliton dynamics [7-10]. In particular, simulations based on the nonlinear Schrödinger equation demonstrate the generation of optical spatial soliton fractals from a single input beam [7] and show optical temporal soliton fractal formation from a single input temporal soliton [8]. This letter reports experimental work that demonstrates for the first time the existence of fractals in soliton dynamics. These "soliton fractals" were produced through the self-generation of spin wave envelope solitons in a magnetic film based active feedback ring system. When the ring gain is increased above the level for stationary single-soliton circulation, the circulating soliton starts to breath in a fractal pattern.

If one amplifies the output signal from a nonlinear dispersive waveguiding medium and then feeds it back to the input of the medium, one forms an active feedback ring system. Such delay-feedback ring systems have proved to be useful in fundamental studies of nonlinear wave dynamics, including envelope solitons [11-15] and chaos [16,17]. Such ring systems also have potential applications for spread spectrum and secure communications [18-20].

In this work, soliton fractals were realized with an active ring system based on a magnetic yttrium iron garnet (YIG) thin film. The film served as the nonlinear dispersive medium for the propagation of spin waves [21]. The propagation geometry was chosen to give an attractive (or self-focusing) nonlinearity that supports the formation of bright spin wave envelope solitons [22,23]. With active feedback present, the spin wave dispersion characteristics of the film are modified significantly because of the high-quality factor resonant response of the ring [24]. This in turn gives rise to the development of soliton fractal structures.

The experimental arrangement is shown in Fig. 1. The YIG film strip is magnetized to saturation by a static magnetic field parallel to the length of the strip. Such a film/field configuration supports the propagation of backward volume spin waves [21] that have an attractive nonlinearity. Two microstrip transducers are placed over the YIG strip to excite and detect the spin wave signals. The output signal from the detection transducer is fed back to the excitation transducer through an adjustable attenuator and a microwave amplifier. The ring signal is sampled through a directional coupler, with feeds to a microwave spectrum analyzer for frequency analysis and a diode detector and a fast oscilloscope for temporal signal display.

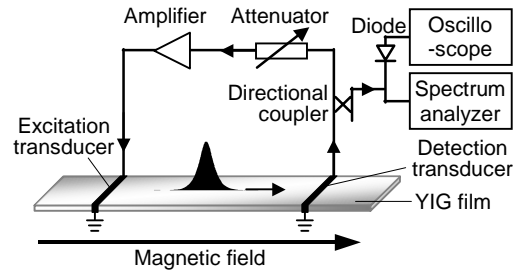

FIG. 1. Diagram of yttrium iron garnet (YIG) film based active ring system. The magnetic field is parallel to the film strip. Microstrip transducers are used for spin wave excitation and detection. The signal from the detection transducer is fed back to the excitation transducer through an attenuator and a microwave amplifier.

For the data given below, the YIG film strip was 6.8 μm thick, 2.2 mm wide, and 46 mm long. It was cut from a larger single crystal YIG film grown on a gadolinium gallium garnet substrate by standard liquid phase epitaxy technique. The film had unpinned surface spins and a nominal 5 GHz ferromagnetic resonance full linewidth in the 0.5 Oe range. The microstrip transducers were 50 μm wide and 2 mm long elements. The transducer separation was held at 7.6 mm. The static magnetic field was set to 1190 Oe. The microwave amplifier had a 30 dB dynamic range, a peak output power of 2 W, and a linear response from 2 to 8 GHz. These characteristics insured that the nonlinear response of the active ring was determined solely by the YIG film.

The active feedback ring system can have a number of resonance eigenmodes that exhibit low decay rates. For a magnetic film active ring, the resonance eigenmode frequencies are determined by the phase condition, $k(\omega) \cdot l + \phi_e = 2\pi n$, where $k$ is the spin wave wavenumber, $\omega$ is the frequency, $l$ is the transducer separation, $\phi_e$ is the phase shift introduced by the electronic loop, and $n$ is an integer. The mode frequencies and the frequency spacing can be adjusted through a change in the $k(\omega)$ dispersion function and/or the transducer separation $l$. The dispersion function, in turn, can be controlled through the film parameters and the applied magnetic field [21]. Note that in a typical active ring soliton self-generation experiment [13,14] about ten eigenmodes are involved. At a low ring gain $G$, all of these eigenmodes experience an overall net loss and there is no spontaneous signal in the ring. If the ring gain is increased to a certain threshold level, here taken as $G = 0$, the eigenmode with the lowest decay rate will start to self-generate and one will obtain a continuous wave response at that eigenmode frequency. Further increases in the ring gain result first in the generation of a comb-like power frequency spectrum, which in the time domain corresponds to a spin wave envelope soliton that circulates in the ring [14], and then in the development of soliton fractal structure that is reported below.

Figure 2 gives a schematic view of the fractal development through a set of simplified power frequency spectrum diagrams for increasing ring gain levels. At the $G = 0$ threshold, the ring eigenmode with the lowest loss appears and the spectrum shows a single frequency peak as in Fig. 2(a). This eigenmode serves as the mother mode or the *initiator* for the fractal development. At some $G = G_1 > 0$, this mother mode produces daughter modes whose frequencies form a comb like spectrum as shown in Fig. 2(b). This set of modes plays the role of the *generator* for the fractal structure. As the gain is then increased to $G_2$, each daughter mode now produces granddaughter modes. One now has a multiple comb frequency spectrum for which each peak in the daughter comb is comprised of a more closely spaced granddaughter comb, as in Fig. 2(c). The process does not stop here. As the gain is increased further to $G_3$, each granddaughter mode produces its own comb of great-granddaughter modes with even more closely spaced frequencies, as in Fig. 2(d). In the language of fractals, the frequency spectra in Fig. 2(c) and (d) are called Prefractal 1 and Prefractal 2, respectively. An infinite sequence of prefractals with an infinitely fine self-similar structure forms *en route* to a fractal in the mathematical sense. In real physical systems, the sequence always terminates at some level.

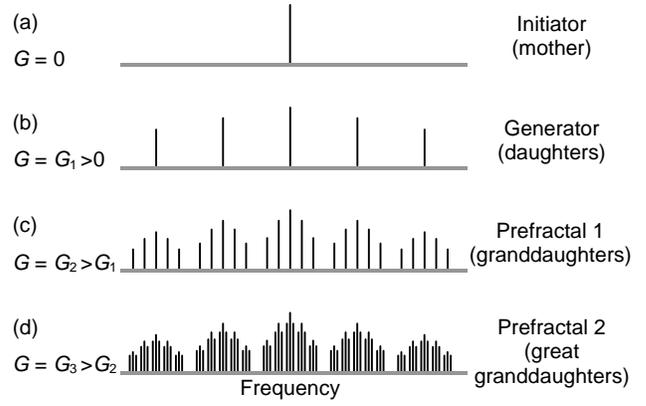

FIG. 2. Schematic power frequency spectra that illustrate fractal development as the ring gain $G$ is increased from zero to some $G_1$, $G_2$, and $G_3$, in sequence.

Figure 3 gives actual YIG film ring experimental spectral data that show fractal development similar to that sketched above. The data in Fig. 3(a), (b), and (c) are for $G$-values of 2.0, 4.6, and 4.7 dB, respectively. The bottom diagrams in the individual panels show the full power frequency spectrum for the ring signal at the indicated gain. The three smaller diagrams in the panels give × 100 expanded views of the three main peaks in the bottom diagrams. All diagrams have the same vertical power scale. The peak widths in all of the diagrams are instrument limited.

In Fig. 3(a), the spectrum for $G = 2.0$ dB shows a daughter frequency comb centered at $f_0 = 5293.23$ MHz. This frequency corresponds to the initial mother eigenmode response obtained at $G = 0$. The daughter comb spacing, $\Delta f_1 = 4.57$ MHz, matches the expected frequency difference between the two neighboring ring eigenmodes with the lowest loss and the ring resonance condition given above. It is important to emphasize that each top small diagram in Fig. 3(a) shows only a single narrow peak. There is no additional structure.

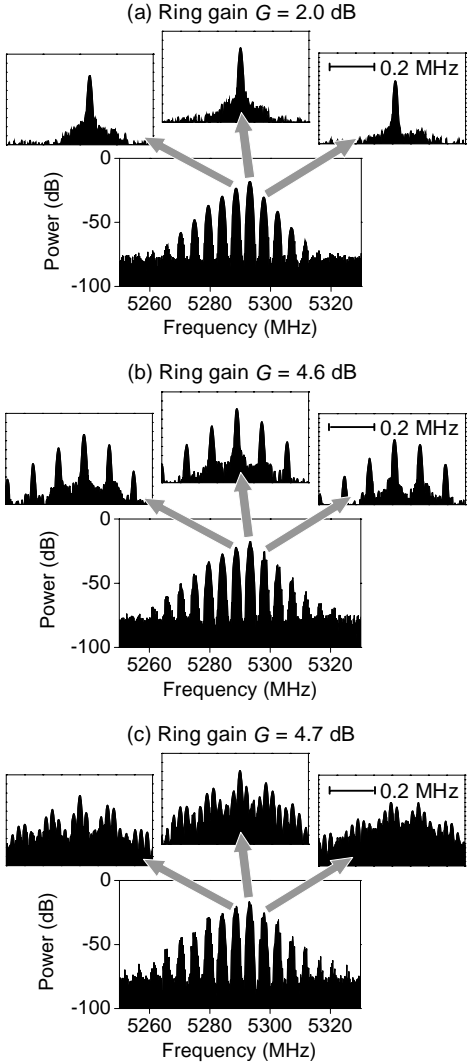

FIG. 3. Power frequency spectra for the magnetic film feedback ring structure at different ring gain $G$-values, as indicated. All vertical power scales are the same. In each panel, the larger diagram shows the full instrument limited comb spectra, and the smaller ×100 expanded frequency scale diagrams show the structure of the three main peaks.

The spectrum in Fig. 3(b) shows the experimental realization of the Prefractal 1 response shown schematically in Fig. 2(c). The instrument-limited spectrum in the bottom diagram is almost the same as in Fig. 3(a). However, the ×100 expanded views now reveal granddaughter combs with a spacing $\Delta f_2 = 0.137$ MHz, much smaller than the daughter comb spacing. Similar granddaughter combs manifested themselves for the majority of the peaks in the daughter comb. The spectrum in Fig. 3(c) takes the data to the next level, with the intriguing experimental realization of a Prefractal 2 response. Now, as one zooms in on the three central daughter comb peaks, the individual granddaughter comb peaks have now generated great-granddaughter combs with an even finer peak spacing at $\Delta f_3 = 0.035$ MHz. All of the great-granddaughter combs have the same spacing. The granddaughter and great-granddaughter spectra in Fig. 3 exhibit clear *self-similarity*.

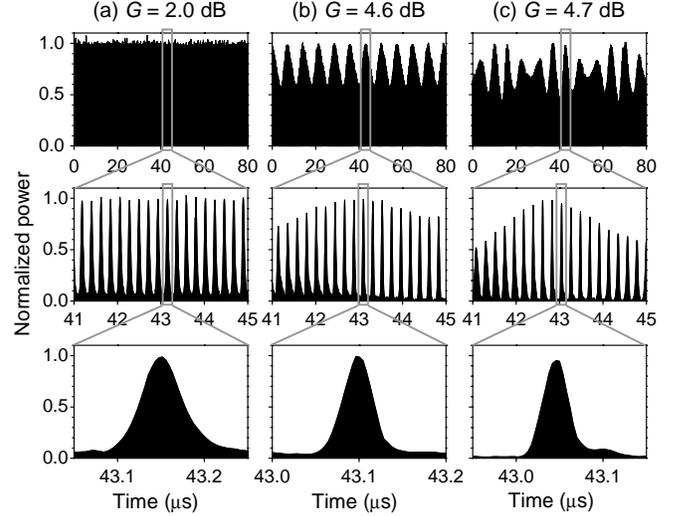

FIG. 4. Sequences of power versus time traces for the YIG film active ring structure at different ring gains, as indicated. From top to bottom, the data in the gray rectangle delineated intervals are shown in ×20 expanded time windows as indicated.

The modal fractal response presented above has extremely important implications for signals in the time domain. Columns (a), (b), and (c) in Fig. 4 show sequences of microwave power versus time profiles that correspond to the power frequency spectra in Fig. 3. The middle row profiles are in a ×20 expanded time window relative to the top row profiles as indicated. The bottom row profiles are in a ×20 expanded time window relative to the middle row profiles as indicated. All vertical power scales are the same.

The data in Fig. 4(a) show the bright soliton train associated with the original low gain (2.0 dB) daughter comb. The soliton train period of 220 ns matches $1/\Delta f_1 = 219$ ns. This period also matches closely to the estimated ring round trip time given by $l/v_g + t_e = 223$ ns, where $v_g$ is spin wave group velocity and $t_e$ is the signal propagation time in the electronic circuits of the ring. Such a match between the train period and the ring round trip time indicates the circulation of a single spin wave envelope soliton in the ring. For the sequence in Fig. 4(b), one sees that the Prefractal 1 granddaughter comb structure in Fig. 3(b) corresponds to an amplitude-modulated or *breathing* soliton train in the time domain. The breathing period of 7.30 μs matches $1/\Delta f_2$. As shown in Fig. 4(c), the time domain signature of the Prefractal 2 great-granddaughter comb structure in Fig. 3(c) is even more interesting. One sees that the soliton breathing modulation in Fig. 4(b) is now itself modulated. The period of this secondary modulation is about 27 μs, close to $1/\Delta f_3$. It is noteworthy that for all three gain cases discussed here, one has a circulating soliton with a fixed period equal to the ring round trip time. This is clearly evident from the middle diagrams in Fig. 4.

In the time domain, one can see, therefore, that the fractal development noted for the signals in the frequency domain carries over directly to a breathing carrier wave in the form of a soliton train, followed by a breathing of the train, and then by a breathing of the breathing. The appearance of wave packets at quite different time scales, as in the Fig. 4(c) diagrams, clearly demonstrates the *self-similarity* in the time domain. The self-similarity can be characterized by a variety of dimension parameters, including a similarity dimension [3,8]. For the data shown in Fig. 4(c), the similarity dimension was estimated to be about 0.8. One can say that the circulating spin wave envelope soliton breathes in fractal patterns. As the ring gain is increased, the frequency-domain fractal structure is built from large scale to small scale, while the time-domain fractal is built from small to large. This reflects the basic nature of the frequency-time Fourier transform. As the gain is increased further, the nonlinear signals generated in the active ring begin to exhibit chaotic behavior, which truncates the sequence of prefractals.

The fractal structure development demonstrated above is a result of changes of the spin wave dispersion properties with the ring gain [24]. Direct dispersion measurements indicate that the dispersion increases significantly as the gain is increased, especially in the vicinity of a ring eigenfrequency. This change in the dispersion, in turn, breaks up the dispersion-nonlinearity balance and leads to the soliton fractal development.

This type of process is similar to the interruption scheme proposed by Segev and co-workers [7,8]. In that scheme, a spatial change in the dispersion or nonlinearity induces the breakup of the soliton balance and produces soliton fractals. The process is also similar to the soliton modulation process reported by Soto-Crespo *et al.* [25]. Here, the circulation of a single optical temporal soliton is excited in a mode-locked fiber ring. An increase in the pumping power or a change in the orientation of mode-locking wave plates modifies the nonlinear loss function of the ring cavity and leads to a soliton modulation response similar to that shown in Fig. 4(b).

In conclusion, this letter reports the first experimental evidence for soliton fractals in a nonlinear dispersive system, and specifically in a magnetic film based feedback active ring. The experiment utilized the self-generation of spin wave envelope solitons in the ring for fractal development. At high ring gain, the temporal soliton circulates in the ring and breathes in a fractal pattern. In the frequency domain, the corresponding spectrum shows a self-similar comb structure.

This work was supported in part by the U. S. Army Research Office, DAAD19-02-1-0197 and W911NF-04-1-0247, the U. S. National Science Foundation, DMR-0108797, the Russian Foundation for Basic Research, 05-02-17714, and the U. S. Department of Energy, Office of Basic Energy Sciences via the Chemical Sciences, Geosciences, and Biosciences Division. Professor M. Segev of the Technion – Israel Institute of Technology is acknowledged for helpful discussions.